\RequirePackage{snapshot}

\documentclass[conference]{IEEEtran}

\usepackage[backend=biber, style=ieee, citestyle=numeric, backref=false]{biblatex}

\DeclareSourcemap{
  \maps[datatype=bibtex]{
    \map[overwrite]{
      \step[fieldsource=doi, final]
      \step[fieldset=url, null]
      \step[fieldset=eprint, null]
    }  	
  }
}

\usepackage{algorithmic}
\usepackage{textcomp}
\usepackage{enumerate}

\usepackage[usenames,dvipsnames,svgnames,table]{xcolor}
\usepackage{graphicx, float, wrapfig, caption, subcaption}

\graphicspath{
}

\usepackage[colorlinks=true, pdfstartview=FitV, linkcolor=Emerald, citecolor=cyan, urlcolor=WildStrawberry]{hyperref}

\usepackage{beesonmacros}

\usepackage{mathabx}

\newcommand{\mN}{\mathcal{N}}

\newcommand{\wt}[1]{\widetilde{#1}}

\newcommand{\mF}{\mathcal{F}}

\newcommand{\orcid}[1]{\href{https://orcid.org/#1}{\includegraphics[scale=.012]{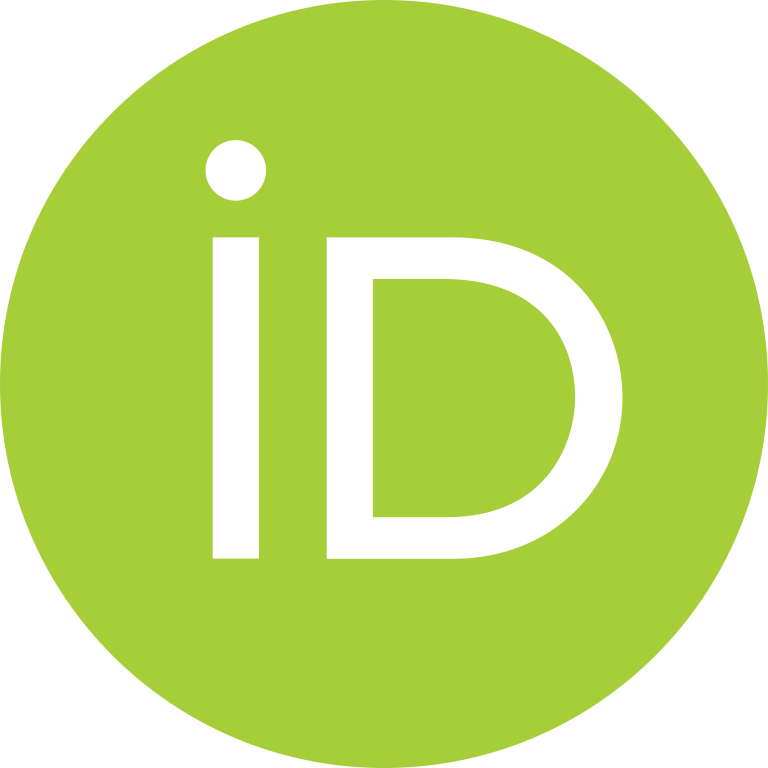}}}

\begin{document}
\begin{refsection}

\title{
Nudged Particle Filter with Optimal Resampling
Applied to the Duffing Oscillator
}

\author{
\IEEEauthorblockN{
Ryne Beeson \orcid{0000-0003-2176-0976}
}
\IEEEauthorblockA{
\textit{Mechanical and Aerospace Engineering} \\
\textit{Princeton University}\\
Princeton, New Jersey, USA \\
ryne@princeton.edu}
\and
\IEEEauthorblockN{
Uwe Hanebeck \orcid{0000-0001-9870-2331}
}
\IEEEauthorblockA{
\textit{Intelligent Sensor-Actuator-Systems Laboratory (ISAS)} \\
\textit{Institute for Anthropomatics and Robotics} \\
\textit{Karlsruhe Institute of Technology (KIT)}\\
Karlsruhe, Germany \\
uwe.hanebeck@kit.edu}
}

\maketitle

\thispagestyle{plain}
\pagestyle{plain}

\begin{abstract}
Efficiently solving the continuous-time signal and discrete-time observation filtering problem for chaotic dynamical systems presents unique challenges in that the advected distribution between observations may encounter a separatrix structure that results in the prior distribution being far from the observation or the distribution may become split into multiple disjoint components. 
In an attempt to sense and overcome these dynamical issues, as well as approximate a non-Gaussian distribution, a nudged particle filtering approach has been introduced. 
In the nudged particle filter method a control term is added, but has the potential drawback of degenerating the 
weights of the particles. 
To counter this issue, we introduce an intermediate resampling approach based on the modified Cram\'{e}r-von Mises distance. 
The new method is applied to a challenging scenario of the non-chaotic, unforced nonlinear Duffing oscillator, which possesses a separatrix structure. 
Our results show that it consistently outperforms the standard particle filter with resampling and original nudged particle filter. 
\end{abstract}


\section{Introduction}
\label{section: introduction}

The extension of particle filtering methods to complex systems, such as those encountered in the geosciences, has not yet been achieved. 
The core issue is the increased occurrence of particle degeneracy in not only high-dimensional systems, but also those that may be chaotic or turbulent, and may receive observations with a temporally sparse cadence. 
Classical studies on the degeneracy of the particle filter in high-dimensional systems include the work by Bengtsson et al. \cite{Bengtsson:2008.ims} and Synder et al. \cite{Snyder:2008mwr}, where simply linear systems with additive Gaussian noise is used. 
The work of Synder et al. \cite{Snyder:2015mwr} has more recently looked at how the optimal proposal provides the best performance bounds to minimize degeneracy in the case of importance sampling particle filters. 

With an aim toward generating samples from the optimal proposal using control theoretic techniques, Lingala et al. \cite{Lingala:2014bu} introduced the nudged particle filter, which solves an optimal control problem to better advect particles toward the optimal proposal, given the future observation. 
We describe this approach in Section \ref{section: standard and nudged particle filter} and build on it to develop a new method described in Section \ref{section: intermediate resampling nudged particle filter}.
A potential drawback of the nudged method is that particles can degenerate during the advection to generate the prior due to excessive control. 
This was better understood in the work of Yeong et al. \cite{Yeong:2020} and studied extensively on the Lorenz 1996 model in Beeson and Namachichivaya \cite{Beeson:2020nd}. 
A simpler linear heuristic control law with the same drawback was presented in van Leeuwen \cite{vanLeeuwen:2010a}. 
Although computationally more expensive than the simple control law of van Leeuwen \cite{vanLeeuwen:2010a}, the nudged particle filter fully senses the nonlinear dynamics in the future forecast, enabling control that is sensitive to dynamical flow with separating behavior. 

Another key advantage of the nudged particle filter for chaotic systems and those especially encountered in the geosciences, is the ability for the conditional distribution to remain near the strange attractor while generating samples approximating the optimal proposal. 
Methods that may move the distribution away from the attractor, such as Kalman approaches or recent flow-based particle methods, then require an advection time after the assimilation update to settle back onto the attractor and be physically realistic. 
The recent work of Zhou and Beeson \cite{Zhou:2024.fusion} looked at the effectivity of projecting flow-based particle methods onto an attractor to maintain this property and Beeson \cite{Beeson:2024.enoc} looked at the value of a hybrid approach using the nudged particle filter with a flow-based particle filter. 
Other approaches and ideas for effective particle filtering on high-dimensional systems have recently been surveyed by van Leeuwen et al. \cite{vanLeeuwen:2019qjrms}.

This paper contributes a new idea to the nudged particle filter by performing an intermediate resampling scheme using the modified Cram\'{e}r-von Mises distance introduced by Hanebeck and Klumpp \cite{Hanebeck:2008.ieee.cmfiis} and Eberhardt et al. \cite{Eberhardt:2010.acc}.  
The aim of this modification to the nudged particle filter is to remedy the degeneracy that may occur due to the control term. 
The original nudged particle filter maintained particle independence during advection between observations, whereas the intermediate resampling naturally results in correlation of the particles. 
Although this is a step back for proving possibly theoretical results, it enables a more continuous control action by the particles. 
We demonstrate this capability on a challenging low-dimensional stochastic nonlinear Duffing oscillator setup in Section \ref{section: numerical experiments}.
The main result is that for a fixed number of particles, the new intermediate resampling nudged particle filter is able to consistently outperform both the standard and nudged particle filters to navigate the filtering distribution across separatrix boundaries and converge toward the true hidden signal. 

We describe the dynamics of the Duffing oscillator test problem in Section \ref{section: the duffing oscillator}, followed by the standard and nudged particle filters in Section \ref{section: standard and nudged particle filter}, the new importance resampling nudged particle filter in Section \ref{section: intermediate resampling nudged particle filter}, the experimental results in Section \ref{section: numerical experiments}, and finally conclusions in Section \ref{section: conclusions}. 

\section{The Duffing Oscillator}
\label{section: the duffing oscillator}

\begin{figure}[t]
\centering
\includegraphics[scale=0.6]{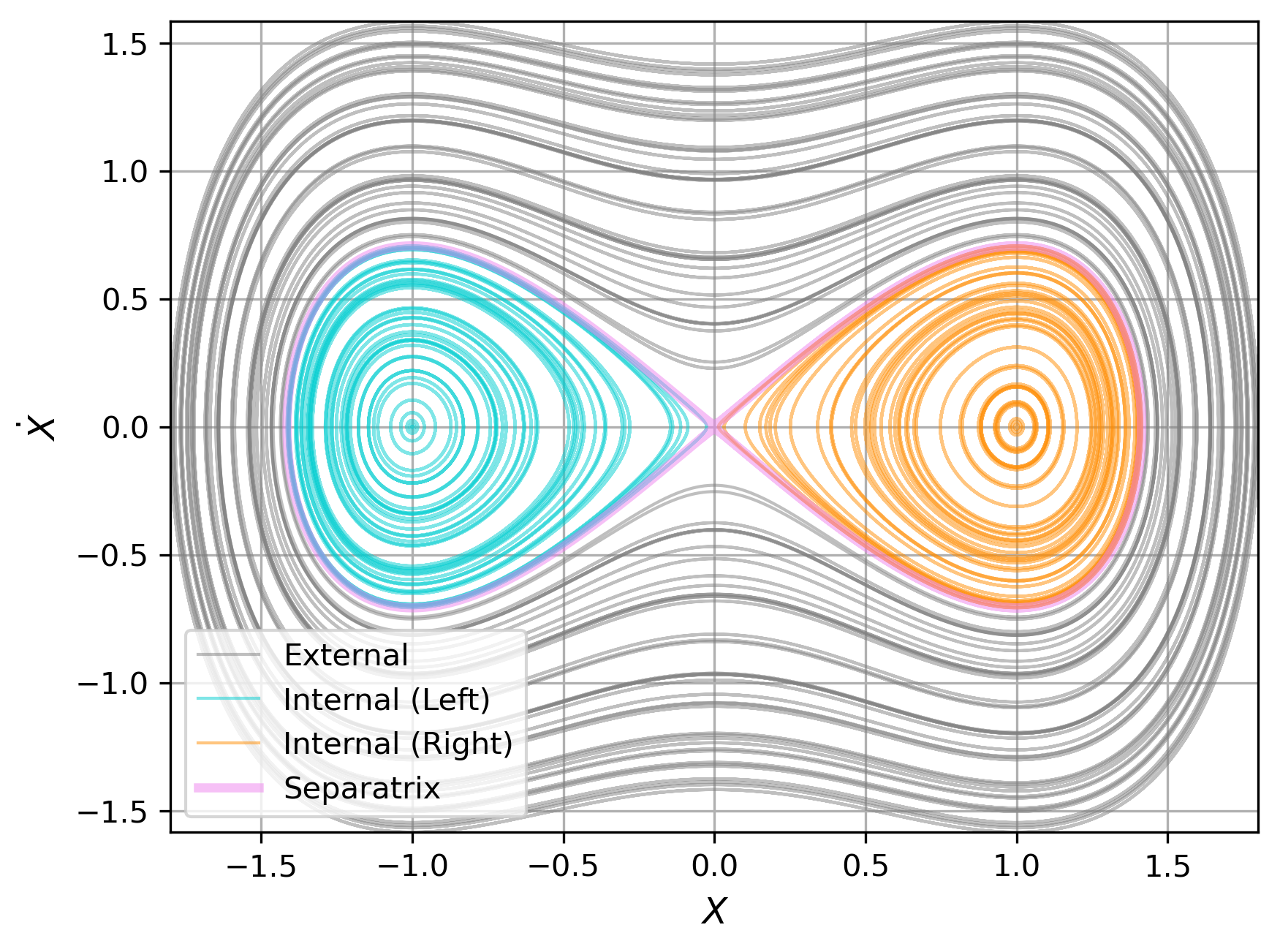}
\caption{Phase portrait for the deterministic unforced nonlinear Duffing oscillator, highlighting realms of dynamical significance.}
\label{figure: the Duffing oscillator}
\end{figure}

The unforced nonlinear Duffing oscillator is a one-degree of freedom Hamiltonian system, which we extend to the stochastic setting by driving it with a Brownian motion. 
The stochastic version is therefore given by the following stochastic differential equation (SDE), 
\begin{align}
\label{equation: nonlinear stochastic Duffing oscillator}
d
\begin{bmatrix} 
x_t \\ \dot{x}_t 
\end{bmatrix}
= \begin{bmatrix} 
0 & 1 \\ 
1 & 0 \end{bmatrix}
\begin{bmatrix} 
x_t \\ \dot{x}_t 
\end{bmatrix} dt 
- 
\begin{bmatrix} 
0 \\ x^3_t
\end{bmatrix} dt 
+ \sigma dW_t,
\end{align}
where $\sigma = \textrm{1E-3} \cdot \op{Id} \in \R{2 \times 2}$, with $\op{Id}$ the identity matrix, and $W$ is a two-dimensional standard Brownian motion (BM). 
The deterministic variant of \eqref{equation: nonlinear stochastic Duffing oscillator} possesses three equilibrium points; one saddle equilibria at the origin, and two center equilibria at $(-1, 0)$ and $(1, 0)$. 
The stable and unstable manifolds of the saddle equilibria generate homoclinic connections that separate the periodic motion about the center equilibria from each other, as well as from a periodic motion that exists in the remaining subset of the phase space. 
These realms of dynamical motion are shown in Fig. \ref{figure: the Duffing oscillator} and the separatrix structure from the saddle equilibria and its invariant manifolds will be exploited in the numerical experiments of Section \ref{section: numerical experiments}. 

\section{Standard and Nudged Particle Filter}
\label{section: standard and nudged particle filter}

For an explanation of the standard particle filter (PF) with resampling, the nudged particle filter (nPF), and the new intermediate resampling nudged particle filter (IRnPF) to be introduced in Section \ref{section: intermediate resampling nudged particle filter}, we consider the following general problem setup for a partially observable continuous-time signal, discrete-time observation system
\EquationAligned{
\label{equation: basic SDE filtering setup}
dX_t 
&= f(X_t) dt + \sigma(X_t) dW_t, &\quad X_0 &= x \in \R{m}, \\
Y_{t_k} 
&= h(X_{t_k}) + \xi_{t_k},  &\quad Y_0 &= 0 \in \R{d}, 
}
where $\xi_{t_k} \sim \mN(0, \Sigma_y)$. 
Standard assumptions on the independence of $X_0 \perp (\xi_{t_k}) \perp W_t$ will be assumed.
For brevity and to improve readability, we will often replace an element of the discrete-time set $(t_k)$ with the corresponding index (e.g., $k$).

The filtering process follows the two recursive steps of: 
1. generating the prior conditional distribution $p_t(x | \mF^Y_{k - 1})$ via the Chapman-Kolmogorov equation
\begin{align*}
p_t(x | \mF^Y_{k - 1}) 
&= \int_{\R{m}} p_{t | k - 1}(x | z) p_k(dz | \mF^Y_{k - 1}), 
\quad t \in [t_{k-1}, t_k),
\end{align*}
followed by; 
2. the update of the posterior conditional distribution $p_t(x | \mF^Y_k)$ by Bayes' formula
\begin{align*}
p_t(x | \mF^Y_k) 
\propto \ p_t(Y_k | x) p_t(x | \mF^Y_{k - 1}), \qquad t = t_k.
\end{align*}

\subsection{Standard Particle Filter with Resampling (PF)}
\label{subsection: standard particle filter with resampling}

The standard particle filter with resampling solves the filtering problem as follows:
\begin{enumerate}[(1)]
\item approximate the initial distribution by a sum of $N$ equally weighted Dirac distributions that have support on the singletons $(X^i_0 \sim p_0)$ which are i.i.d., 
\begin{align*}
p_0^N(dx | Y_0) 
\equiv \sum_{i = 1}^N w^i_0 \delta_{X^i_0}(dx) 
= \frac{1}{N} \sum_{i = 1}^N \delta_{X^i_0}(dx);
\end{align*}
\item at time $k-1$, the prior is generated under the action of the Markov kernel $P_X$ for the dynamics in \eqref{equation: basic SDE filtering setup}, 
\begin{align*}
p_k^N(dx | \mF^Y_{k - 1})
&= p^N_{k - 1}(dx | \mF^Y_{k - 1}) P_X  
= \sum_{i = 1}^N w^i_{k-1} \delta_{X^i_k}(dx);
\end{align*}
\item using the observation $Y_k$, the unnormalized weights are calculated as
\begin{align}
\label{equation: unnormalized weights, standard particle filter}
\wt{w}^i_k 
= w^i_{k - 1} p_k(Y_k | X^i_k);
\end{align}
\item normalize the particle weights, 
\begin{align*}
w^i_k = \wt{w}^i_k / \sum_{i = 1}^N \wt{w}^i_k;
\end{align*}
\item if the particle weights are degenerate
, resample the particles and set all weights to be equal (i.e., $w^i_k = 1 / N$); 
\item define the posterior at time $k$ as
\begin{align*}
p^N_k (dx | \mF^Y_k)
\equiv
\sum_{i = 1}^N w^i_k \delta_{X^i_k} (dx),
\end{align*}
 and return to Step (2).
\end{enumerate}
In this work, we use an approximation to the effective sample size, 
which was introduced in Bergman \cite{Bergman:1999} and Liu and Chen \cite{Liu:1998jasa}, as the metric for Step (5) in deciding whether degeneracy has occurred. 
Denoting the vector of normalized weights as $w$, the approximate effective sample size is defined as 
\begin{align*}
N_{\textrm{eff}}
\equiv 1 / \langle w, w \rangle.
\end{align*}
We use the threshold of $N_{\textrm{eff}} < N / 2$ to trigger resampling, and use the universal (also known as systematic) resampling approach of Kitagawa \cite{Kitagawa:1996jcgs}. 

\subsection{Nudged Particle Filter (nPF)}
\label{subsection: nudged particle filter}

As described in Section \ref{section: introduction}, the nudged particle filter (nPF) aims to mitigate PF degeneracy in the chaotic and sparse temporal observation case by solving an optimal control problem that nudges particles independently toward the optimal proposal given the future observation. 
Hence, the nPF follows the same steps as the PF, but replaces the advection of particles in Step (2) of Section \ref{subsection: standard particle filter with resampling} with the following dynamical law 
\begin{align}
\label{equation: SDE for signal process in optimal control problem}
dX_t 
= f(X_t) dt + u_t dt + \sigma(X_t) dW_t, \quad t \in [t_k, t_{k + 1}),
\end{align}
where the control term $u$ is the minimizer of the finite-horizon objective function
\begin{multline*}
J(u; k, x; Y_{k + 1}) \\
= \E{ \int_{t_k}^{t_{k + 1}} \frac{1}{2} \langle u_s, R_s u_s \rangle ds + g(X^{k, x}_{k + 1}, Y_{k + 1}) }.
\end{multline*}
In this objective function, $R$ takes values in the space of symmetric positive definite matrices and $g$ is a terminal cost function that penalizes the realization $X^{k, x}_{k + 1}$ from being far from the observation $Y_{k + 1}$. 
In this work, we use the negative log of the observation likelihood for $g$. 
The superscript for $X^{k, x}_{k + 1}$ indicates a realization of \eqref{equation: SDE for signal process in optimal control problem} starting at the state $x \in \R{m}$ at time $t_k$ and advected to time $t_{k + 1}$. 

The solution to this optimal control problem is given by the Hamilton-Jacobi-Bellman equation, which has the optimal feedback control law solution
\begin{align}
\label{equation: feedback control solution of HJB}
u(t, x) 
= - R^{-1}_t \nabla_x V(t, x),
\end{align}
where $\nabla_x V$ is the gradient of the value function. 
By using a log-transformation \cite{Fleming:1982}, $V(t, x) = -\log \Phi(t, x)$, the control solution takes the form
\begin{align*}
u(t, x) = \frac{1}{\Phi(t, x)} R^{-1}_t \nabla_x \Phi(t, x).
\end{align*}
Assuming the dispersion coefficient $\sigma$, is constant and independent of $X_t$, and then choosing $R^{-1} = \sigma \sigma^*$, the evolution equations of $\Phi$ and $\nabla_x \Phi$ both become backward linear second order parabolic partial differential equations. 
Therefore they have solutions given by the Feynman-Kac formulas 
\begin{align}
\label{equation: Feynman-Kac application for log-transformation solution}
\Phi(t, x) 
= \E{ \exp( - g(\eta^{k, x}_{k + 1}, Y_{k + 1})) },  
\end{align}
and
\begin{multline}
\label{equation: Feynman-Kac application for log-transformation solution of derivative}
\nabla_x \Phi(t, x) 
= -\mathbb{E} \left[ \exp( - g(\eta^{k, x}_{k + 1}, Y_{k + 1})) \right. \\
\left. \exp \left( \int_{t_k}^{t_{k + 1}} \nabla_x f(\eta^{k, x}_s) ds \right) \nabla_x g(\eta^{k, x}_{k + 1}, Y_{k +1}) \right],
\end{multline}
where $\eta$ is a realization to an SDE with the same generator as the signal process in \eqref{equation: basic SDE filtering setup}. 

As alluded to in Section \ref{section: introduction}, the drawback of the nPF is that the weights of the particles must be adjusted if control is performed. 
This is due to the fact that the controlled dynamics \eqref{equation: SDE for signal process in optimal control problem} are not the same as the original signal dynamics \eqref{equation: basic SDE filtering setup}.
The change in weight for a particle $X^i$ is therefore given by a Radon-Nikodym derivative
\begin{multline}
\label{equation: radon-nikodym derivative correction}
\frac{d\mu^i}{d\widehat{\mu}^i}
= \exp \left( -\int_{t_k}^{t_{k + 1}} \langle v(s, X^i_s) , dW_s \rangle \right. \\
\left. - \frac{1}{2} \int_{t_k}^{t_{k + 1}} \langle v(s, X^i_s) , v(s, X^i_s) \rangle ds \right),
\end{multline}
where $v(s, X^i_s) = - \sigma^* \nabla_x V(s, X^i_s)$, and the derivation follows from the results on Girsanov transformations. 
In contrast to \eqref{equation: unnormalized weights, standard particle filter}, the unnormalized weights of the nPF are 
\begin{align}
\label{equation: unnormalized weights, nudged particle filter}
\wt{w}^i_k = \frac{d \mu^i}{d \widehat{\mu}^i} w^i_{k - 1} p_k(Y_k | X^i_k). 
\end{align}
We direct the reader to Yeong et al. \cite{Yeong:2020} for more details on the derivation and properties of the nPF. 

The control solution for the nPF is continuous in time and requires the evaluation of the expectations in \eqref{equation: Feynman-Kac application for log-transformation solution} and \eqref{equation: Feynman-Kac application for log-transformation solution of derivative}. 
An approximation must be made in both of these cases for numerical implementation. 
We therefore do the following in this work: 
\begin{enumerate}[(2a)]
\item
we discretize the time interval $[t_k, t_{k + 1})$ into a uniform partition of $M$ subintervals defined by the collection of times $(t_{k(j)})_{j = 1}^M$;
\item
at the start of each of these intervals, a control solution satisfying \eqref{equation: feedback control solution of HJB} is computed using $K$ realizations;
\item
the control solution is fixed for the control step interval $[t_{k(j)}, t_{k(j + 1)})$, and the particle is advected under the dynamics of \eqref{equation: SDE for signal process in optimal control problem};
\item 
the correction to the unnormalized weight is made according to \eqref{equation: radon-nikodym derivative correction};
\item 
if the particle weight did not degenerate below some threshold, return to Step (2a) to solve for the next control step until the final time $t_{k+1}$ is reached; 
otherwise if the particle did degenerate due to the control solution, advect the particle from $t_{k(j)}$ until $t_{k(j + 1)}$ under the natural dynamics of \eqref{equation: basic SDE filtering setup} without control. 
\end{enumerate}

\section{Intermediate Resampling Nudged Particle Filter (IRnPF)}
\label{section: intermediate resampling nudged particle filter}

Each particle of the nPF calculates their control independently of all others and therefore the unnormalized weights during the intermediate time between observations can become degenerate relative to the original distribution's mean. 
A key contribution in this work is to introduce a deterministic optimal resampling after each control step, such that the advected distribution will have minimum particle weight variance throughout the advection (i.e., $w^i = 1 / N$). 
The benefits of performing this resampling are at least twofold. 
On a local level, each particle has a weight that is reset to $1 / N$, and therefore enables the ability to apply control in Step (2c) of the nPF algorithm without running into the degeneracy condition specified in Step (2e) during the next control step (i.e., any degeneration is isolated to a given control step). 
The new IRnPF is therefore able to continually apply control throughout the intermediate time to drive the particle toward the observation neighborhood based on the terminal constraint $g$. 
On a global level, the resampling enables the use of indirect information generated by each particle on the difficulty of the dynamics to be encountered, so that particles are resampled at locations with the expectation that they will have similar control efforts for the next control step, and hence a similar decrease in their unnormalized weights.

The resampling of particles is achieved by minimizing the modified Cram\'{e}r-von Mises distance, which is defined in Definition \ref{definition: modified Cramer-von Mises distance}. 
This distance was introduced by Hanebeck and Klumpp \cite{Hanebeck:2008.ieee.cmfiis} and allows for the comparison of two probability distributions. 
The distance relies on the following definition for the localized cumulative distribution (LCD). 


{
\begin{definition}[Localized Cumulative Distribution \cite{Hanebeck:2008.ieee.cmfiis}]
Given a probability distribution $\mu$, a width parameter $b \in \R{n}_+$, and kernel function $\mathcal{K} : \R{n} \times \R{n} \rightarrow [0, 1]$, the corresponding Localized Cumulative Distribution (LCD) is defined as
\begin{align*}
F_\mu(\cdot ; b, \mathcal{K}) : \R{n} &\longrightarrow [0, \infty), \\
(m) &\longmapsto \int_{\R{n}} \mathcal{K}(x - m, b) \mu(dx). 
\end{align*}
\end{definition}
}

We assume that the kernel function is a separable Gaussian kernel, dependent on a scalar width in each directional component, 
\begin{align}
\label{equation: separable Gaussian kernel}
\mathcal{K}_b(x) \equiv \mathcal{K}(x; b) = \exp \left( - \| x \|^2_2 / 2 b^2 \right), 
\end{align}
which then gives the modified Cram\'{e}r-von Mises distance defined in the following manner:

{
\begin{definition}[Modified Cram\'{e}r-von Mises Distance \cite{Hanebeck:2008.ieee.cmfiis}]
\label{definition: modified Cramer-von Mises distance}
Let $F_\mu$ and $F_\nu$ be the LCD for two probability distributions $\mu$ and $\nu$, both with kernel $\mathcal{K}$, and $w : \R{}_+ \rightarrow \R{}_+$ a weighting function. 
Then their modified Cram\'{e}r-von Mises distance is defined as 
\begin{align*}
D(\mu, \nu)
\equiv \int_\R{} w(b) \int_{\R{n}} \left( F_\mu(m; b, \mathcal{K}) - F_\nu(m; b, \mathcal{K}) \right)^2 dm db.
\end{align*}
\end{definition}
}

In this work, we make use of an asymptotic approximation for $D(\mu, \nu)$ between two Dirac mixtures as $b \rightarrow \infty$ and the weight function is restricted to $(0, \overbar{b}]$ and chosen according to $w(b) = 1 / b^{n - 1}$ by Eberhardt et al. \cite{Eberhardt:2010.acc} (see also Hanebeck \cite{Hanebeck:2015.at.63.4}) in the following theorem. 

{
\begin{theorem}
With $\overbar{b} \gg 1$, $w(b) = 1 / b^{n - 1}$, $\mu, \nu$ Dirac mixtures with support $(x_i), (y_i)$ and weights $(w^i_x), (w^i_y)$ respectively, and $\mathcal{K}$ given according to \eqref{equation: separable Gaussian kernel}, $D(\mu, \nu)$ is approximately,
\begin{multline}
\label{equation: modified Cramer-von Mises distance}
D(\mu, \nu) 
\approx 
\frac{\pi^{n / 2}}{8} \left( \wt{D}(\mu) - 2 \wt{D}(\mu, \nu) + \wt{D}(\nu) \right) \\
+ \frac{\pi^{n / 2}}{4} C(\overbar{b}) \overbar{D}(\mu, \nu), 
\end{multline}
where
\begin{align*}
\wt{D}(\mu, \nu) 
&\equiv \sum_{i = 1}^N \sum_{j = 1}^K w^i_x w^j_y \op{xlog} \left( \| x_i - y_j \|^2_2 \right) \\
\wt{D}(\mu) 
&\equiv \wt{D}(\mu, \mu), 
\quad \wt{D}(\nu) \equiv \wt{D}(\nu, \nu),   \\
\overbar{D}(\mu, \nu) 
&\equiv \| \E[\mu]{x} - \E[\nu]{x} \|^2_2, \\
C(\overbar{b})
&\equiv \log (4 \overbar{b}^2) - \Gamma, 
\end{align*}
where $\Gamma \approx 0.5772$ is the Euler gamma constant and $\op{xlog}(s) \equiv s \log(s)$. 
\end{theorem}
}

\subsection{The IRnPF Method}
\label{subsection: the IRnPF method}

The IRnPF method, an extension of the nPF in Section \ref{subsection: nudged particle filter} is now described. 
In particular, following Step (2e) of the nPF, we have the additional steps: 
\begin{enumerate}[(2a)]
\setcounter{enumi}{5}
\item
compute the normalized weights for the controlled Dirac mixture $p^N_{k(j)}$ and generate a new mixture $\wt{p}^K_{k(j)}$ with $\gamma K = N$\footnote{For the IRnPF we specify $\gamma, K$ first and then $N$ before applying the method, so that this relation holds.}, $\gamma \in \N{}$, evenly weighted samples of the normalized distribution $p^N_{k(j)}$;
\item
translate the samples of $\wt{p}^K_{k(j)}$ to have the same mean as $p^N_{k(j)}$;
\item
using the support locations of $\wt{p}^K_{k(j)}$ as optimization parameters, minimize the modified Cram\'{e}r-von Mises distance between $p^N_{k(j)}$ and $\wt{p}^K_{k(j)}$;
\item 
using the optimized support locations, multiple each particle $\gamma$-times and normalize the weights by $1/\gamma$ to produce the resampled distribution $p^N_{k(j)}$; return to Step (2a) of the nPF. 
\end{enumerate}
Note that the translation given in Step (2g) implies that $\overbar{D} = 0$ in \eqref{equation: modified Cramer-von Mises distance} for the initial iteration of the optimization. 
In this work we do not apply the intermediate resampling after the last control step, since the universal resampling algorithm will be applied after Bayes' update if needed. 

\section{Numerical Experiments}
\label{section: numerical experiments}

In the experiments to follow, we use the following discrete-time observation process,
\begin{align*}
Y_{t_k} 
= 
\begin{bmatrix} 
x_{t_k} \\ \dot{x}_{t_k} 
\end{bmatrix} 
+ \xi_{t_k},
\qquad
\xi_{t_k} \sim \mN(0,  \textrm{1E-2} \cdot \op{Id}),
\end{align*}
with observations occurring every 0.5 time-units (TU) (i.e., $t_{k + 1} - t_k = 0.5$ for all $k$), and $(\xi_{t_k}) \perp X_0 \perp (W_t)$. 
In all simulations, we use a duration of 4.5 TU, and therefore 9 observations. 
All methods make use of the universal resampling approach if the effective sample size falls below 5 following the posterior update (Step (5) of the PF). 
An RK4Maruyama numerical integration scheme with a stepsize of 1E-2 is used for advection of particles.

\begin{figure}[t]
\centering
\includegraphics[scale=0.54]{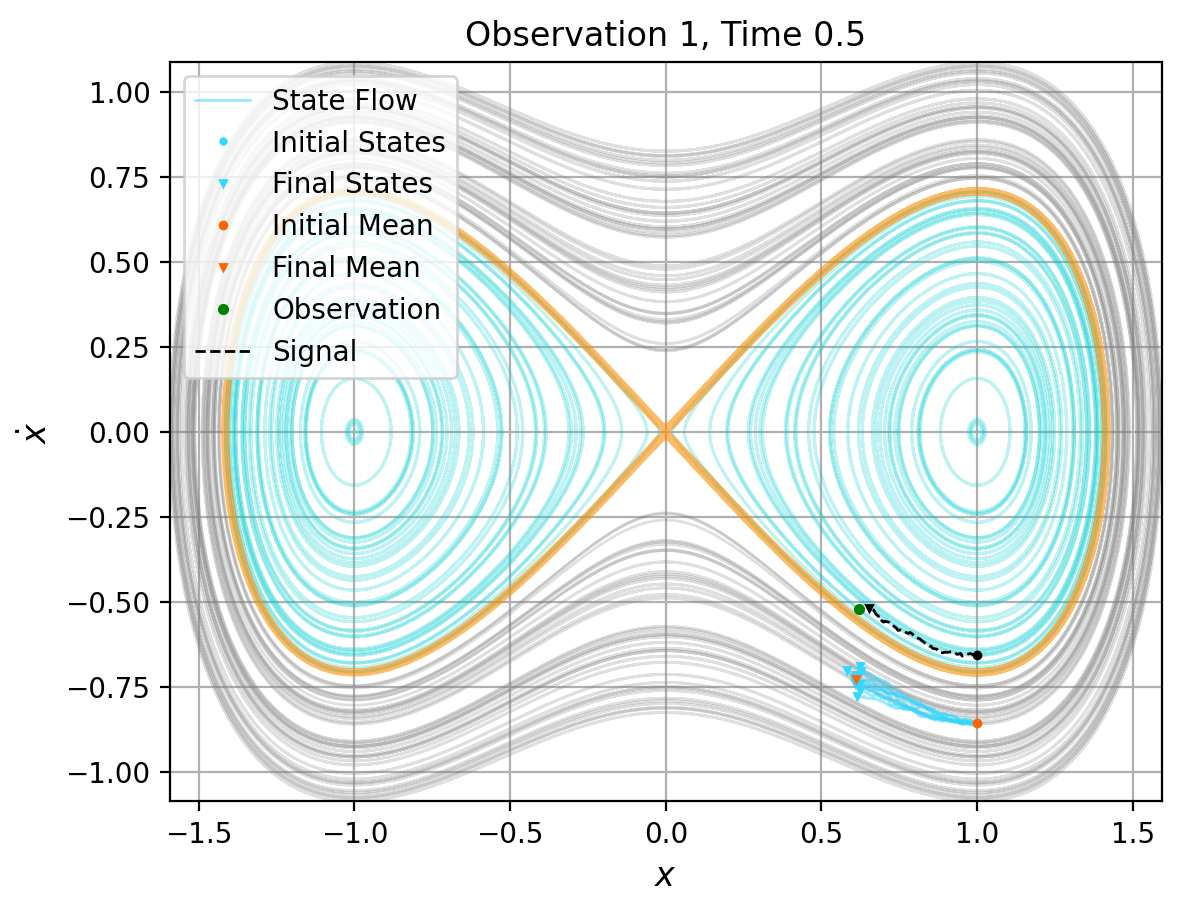}
\caption{
Global view for a realization of the advected initial PF distribution, started with a deterministic initial condition and separated from the true hidden signal state, to the first observation time.
}
\label{figure: global view, separated by stable manifold, prediction flow 0, PF}
\end{figure}

\begin{figure}[t]
\centering
\includegraphics[scale=0.54]{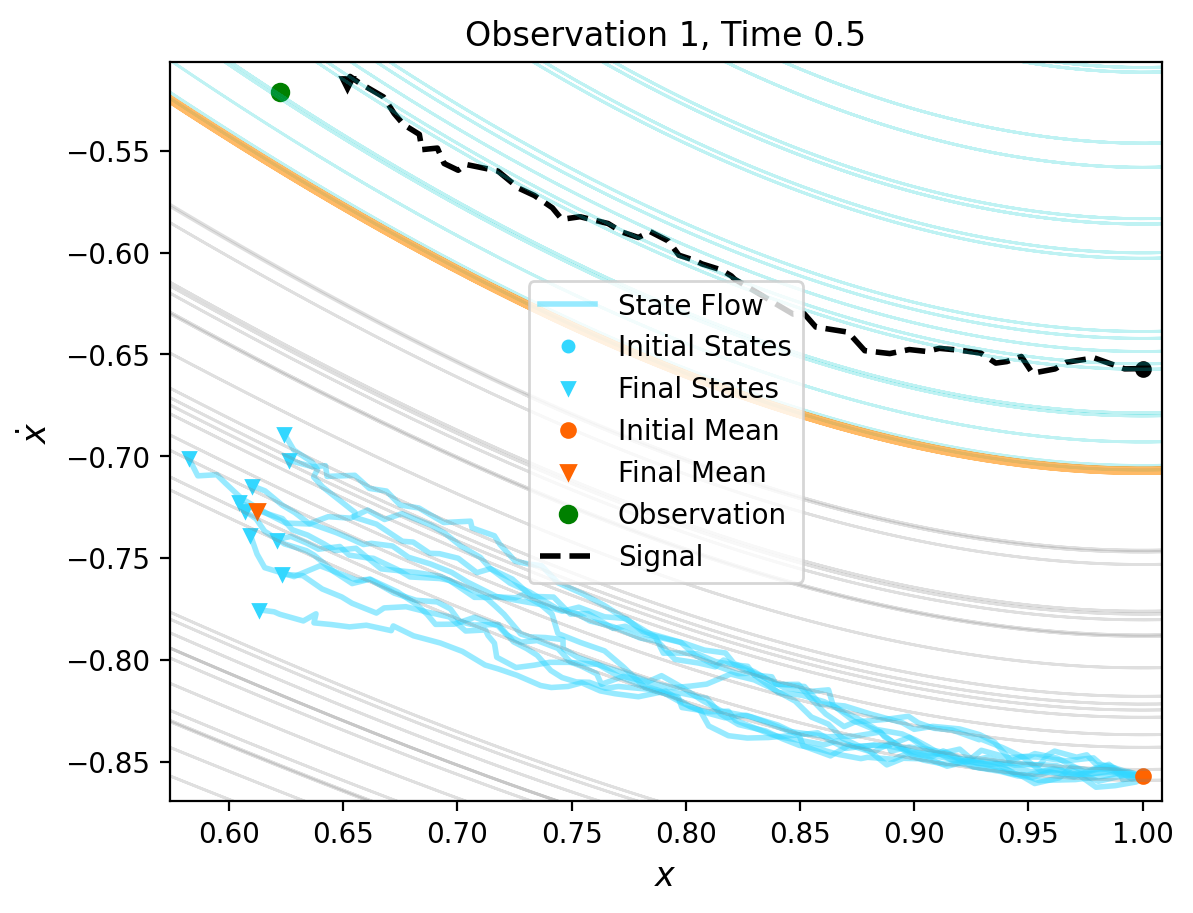}
\caption{
Local view for a realization of the advected initial PF distribution, started with a deterministic initial condition and separated from the true hidden signal state, to the first observation time.
}
\label{figure: local view, separated by stable manifold, prediction flow 0, PF}
\end{figure}

Each particle filtering method uses 10 particles, and $\gamma = 5$ for the IRnPF. 
Therefore $N = 10$ for the PF and nPF, and for the IRnPF we have $\gamma = 5, K = 10$ and hence $N = 50$. 
Note that the IRnPF will have a computationally complexity on the same order as the nPF since the control is still only calculated for 10 particles for any of the control steps. 
The advection using 5 realizations for each control in the IRnPF enables a better approximation of the advected distribution, and hence improved resampling.  
The optimization at Step (2h) for the IRnPF is accomplished using the \textit{scipy.minimize} BFGS method with a (relative) tolerance of 1E-3. 

To demonstrate the qualitative behavior of the three methods, we choose deterministic initial conditions for the signal state at $(x, \dot{x}) = (1, -0.657)$ and the initial filtering distribution a Dirac distribution at $(x, \dot{x}) = (1, -0.857)$. 
This choice results in the signal state being separated from the initial distribution by the deterministic stable manifold (separatrix) to the saddle equilibria. 
These initial conditions are shown in Fig. \ref{figure: global view, separated by stable manifold, prediction flow 0, PF} and \ref{figure: local view, separated by stable manifold, prediction flow 0, PF}, which provides both a global and local (zoomed-in) view of the first observation interval for a PF simulation. 
The separatrix structure with invariant manifolds is shown in orange in these figures. 
The dispersion coefficient in \eqref{equation: basic SDE filtering setup} is such that the deterministic dynamics are not overpowered by the diffusion of the BM, and therefore forward in time, there is a high likelihood that the signal will remain in the \emph{Internal (Right)} realm show in Fig. \ref{figure: the Duffing oscillator}, the filtering distribution will remain in the \emph{External} realm, and the two will diverge as they approach the saddle equilibria. 
The last observation period for this realization of the PF is shown in Fig. \ref{figure: global view, separated by stable manifold, prediction flow 8, PF} and confirms that this is the typical behavior that occurs. 

In Fig. \ref{figure: global view, separated by stable manifold, prediction flow 8, nPF} we show the final observation interval for the same simulation, but using the nPF. 
In particular, this simulation has the same initial conditions, the same signal path, and the same observations of the signal. 
The time between observations (0.5 TU) is partitioned into 50 equal control steps for the nPF, and $K=10$ realizations are used for each particle to calculate their control term in Step (2b). 
Comparing the final observation interval of Fig. \ref{figure: global view, separated by stable manifold, prediction flow 8, nPF} for the nPF to Fig. \ref{figure: global view, separated by stable manifold, prediction flow 8, PF} for the PF is informative; it shows the nPF is lagging behind the PF in the natural dynamical flow. 
This implies that the nPF has been attempting to overcome the dynamical flow to move in the direction of the signal observation. 
Figure \ref{figure: deterministic initial condition, split by stable manifold, particle weight history, nPF} further demonstrates this is the case by viewing the particle paths in the $\dot{x}$ state over the simulation time, and coloring the particle paths according to their relative unnormalized weight during advection. 
Particles that are shaded blue is an indication that they have been actively using control to nudge themselves. 
As indicated in Step (2e) of the nPF algorithm, the particles are only allowed to nudge so much as to not fully degenerate and therefore control is only being applied for a fraction of the total simulation time. 

Figure \ref{figure: global view, separated by stable manifold, prediction flow 8, IRnPF} shows the final observation interval of the IRnPF for this same case. 
In contrast to the nPF, the IRnPF is able to apply control more consistently throughout the simulation due to the intermediate resampling, and therefore is able to navigate the filtering distribution over the separatrix structure and approach the location and observations of the signal process. 
The improvement in the relative unnormalized weights of the particles is shown in Fig. \ref{figure: deterministic initial condition, split by stable manifold, particle weight history, IRnPF}. 
The final control steps of each observation interval shows some particles in blue, which is expected because control is being used and as we stated after Step (2i) for the IRnPF algorithm, an intermediate resampling is not used on the final control step before an observation. 

\subsection{Monte Carlo Results}
\label{subsection: separated by the stable manifold, monte carlo}

A Monte Carlo (MC) analysis was performed to quantify and confirm the typical behavior of the PF, nPF, and IRnPF methods on the experimental setup just described.
In the MC analysis, we use 20 different signal and observation history pairs, and for each pair we perform 20 different simulations. 
Therefore, a total of 400 simulations are produced for each method. 
To be clear, the benchmark of each method is on the same signal and observation pairs, so that direct comparison of performance is possible. 

\begin{figure}[hp]
\centering
\includegraphics[scale=0.54]{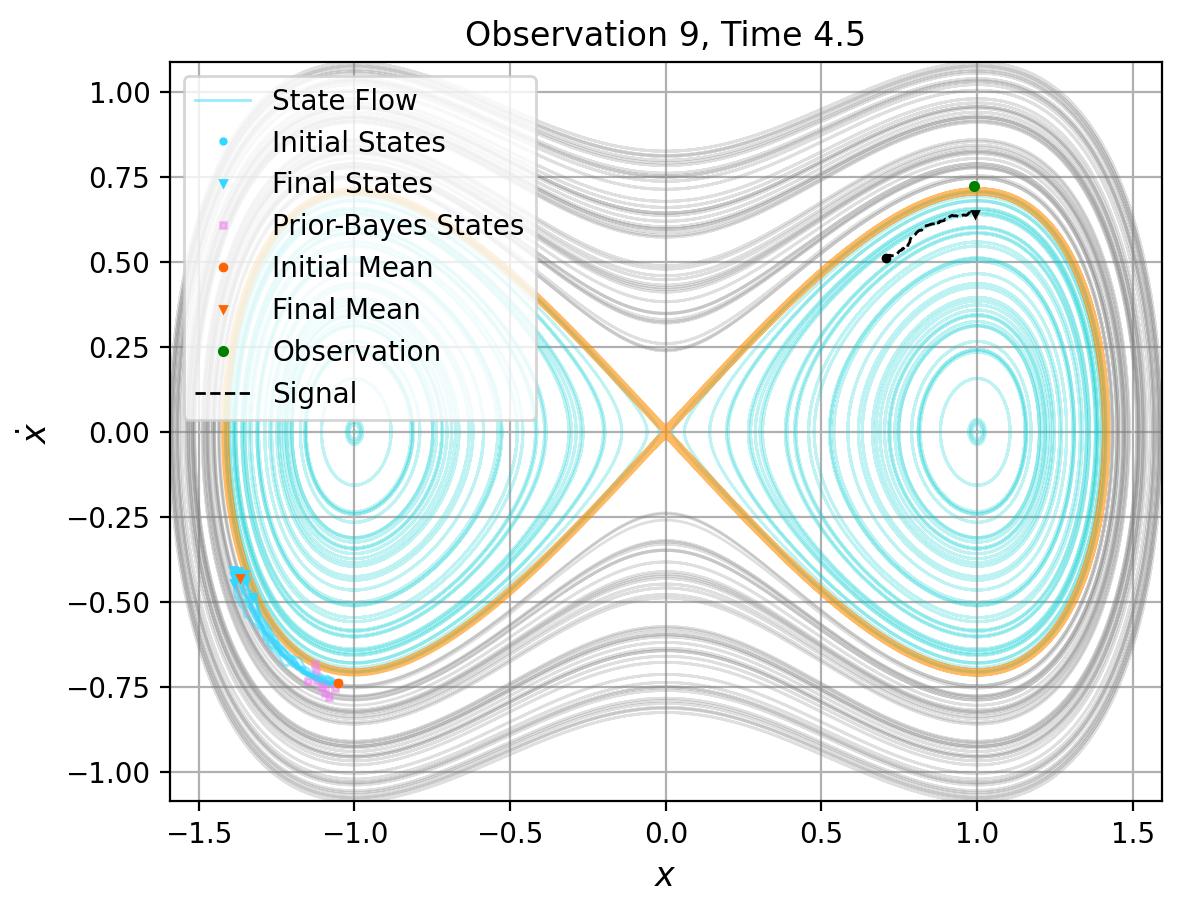}
\caption{
Global view of the PF behavior from the second to last observation to the last observation of the simulation. 
}
\label{figure: global view, separated by stable manifold, prediction flow 8, PF}
\end{figure}

\begin{figure}[hp]
\centering
\includegraphics[scale=0.54]{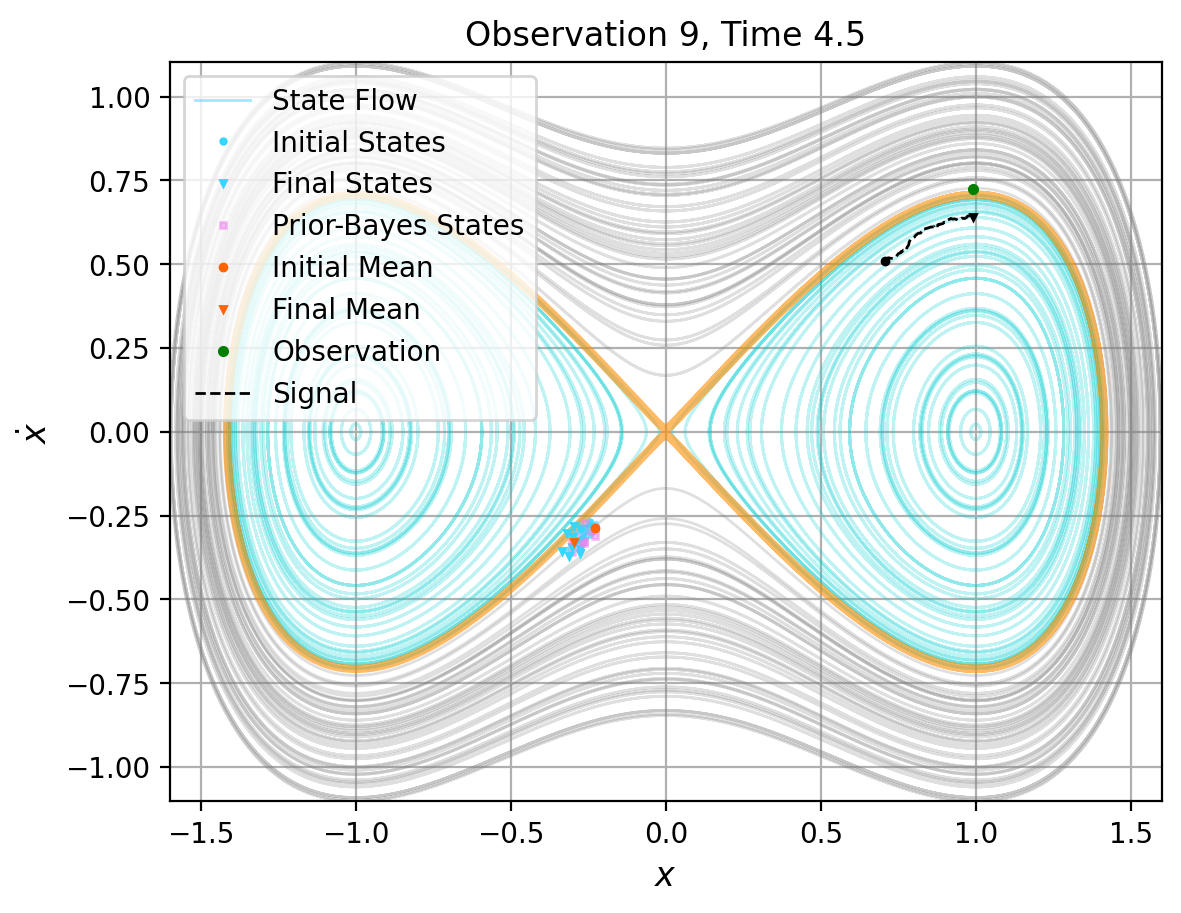}
\caption{
Same as Fig. \ref{figure: global view, separated by stable manifold, prediction flow 8, PF}, but for the nPF. 
}
\label{figure: global view, separated by stable manifold, prediction flow 8, nPF}
\end{figure}

\begin{figure}[hp]
\centering
\includegraphics[scale=0.54]{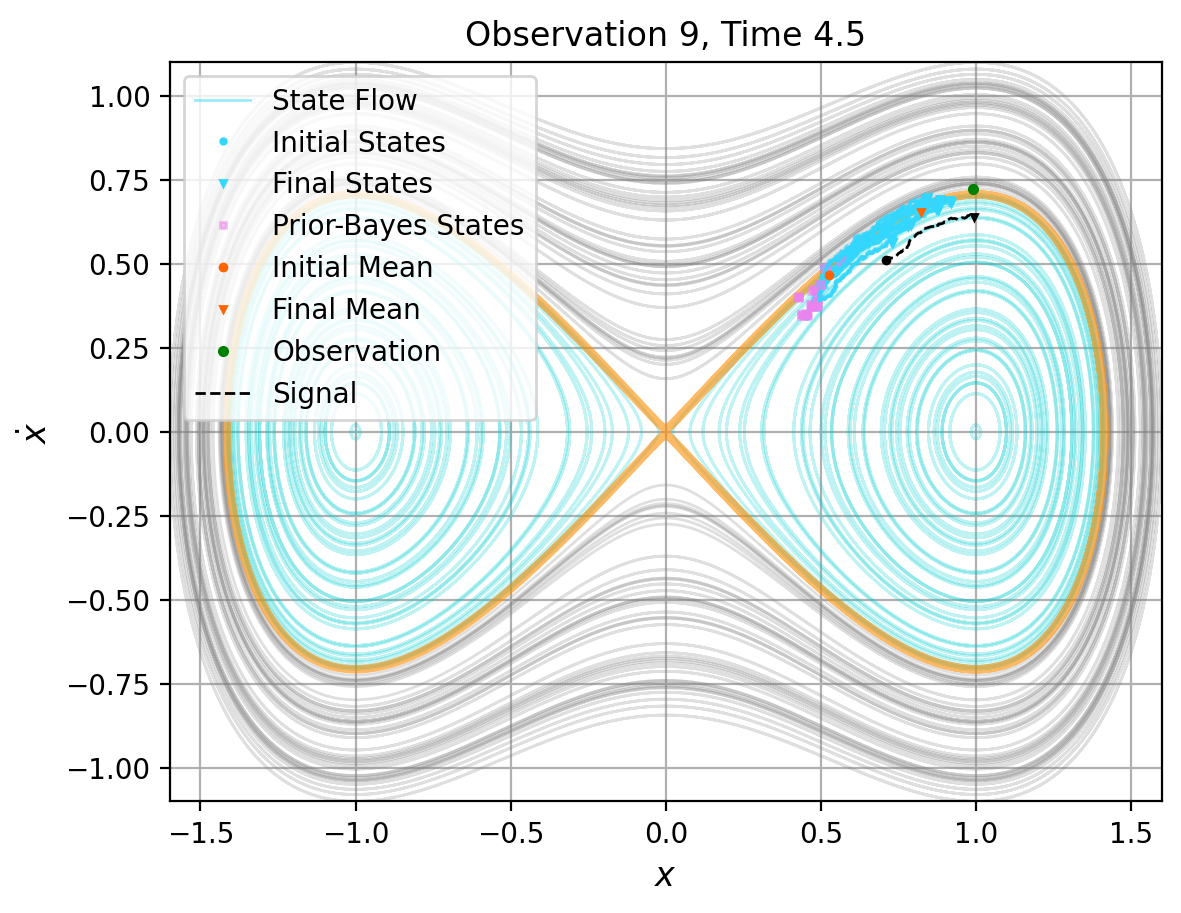}
\caption{
Same as Fig. \ref{figure: global view, separated by stable manifold, prediction flow 8, PF}, but for the IRnPF. 
}
\label{figure: global view, separated by stable manifold, prediction flow 8, IRnPF}
\end{figure}

\begin{figure}[hp]
\centering
\includegraphics[scale=0.6]{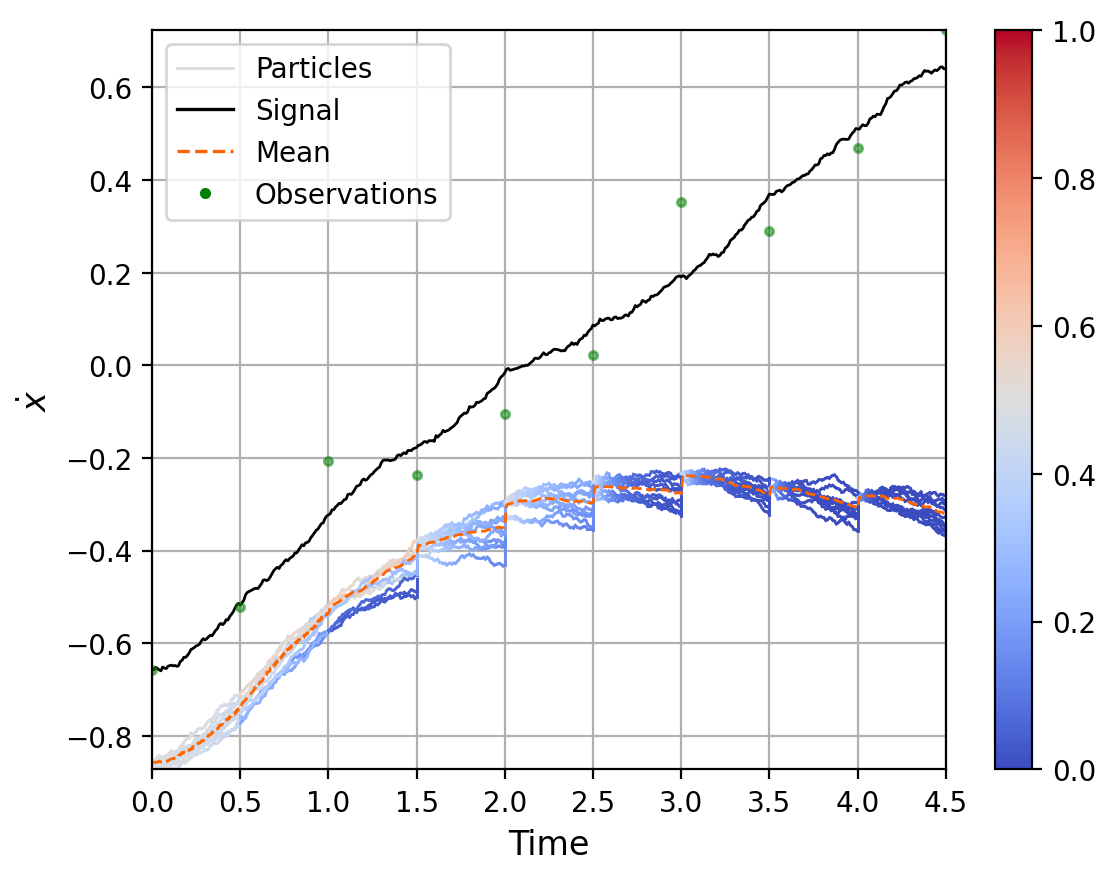}
\caption{
The $\dot{x}$ states versus time for particles of the nPF, with particle paths colored according to their relative unnormalized weights during advection. 
The figure illustrates the disadvantage of possible decreasing unnormalized weights during advection for the nPF; low relative weight (i.e., decreased relative to the mean) is shown in blue, even relative weight in ``white", and high relative weight in red. 
}
\label{figure: deterministic initial condition, split by stable manifold, particle weight history, nPF}
\end{figure}

\begin{figure}[hp]
\centering
\includegraphics[scale=0.6]{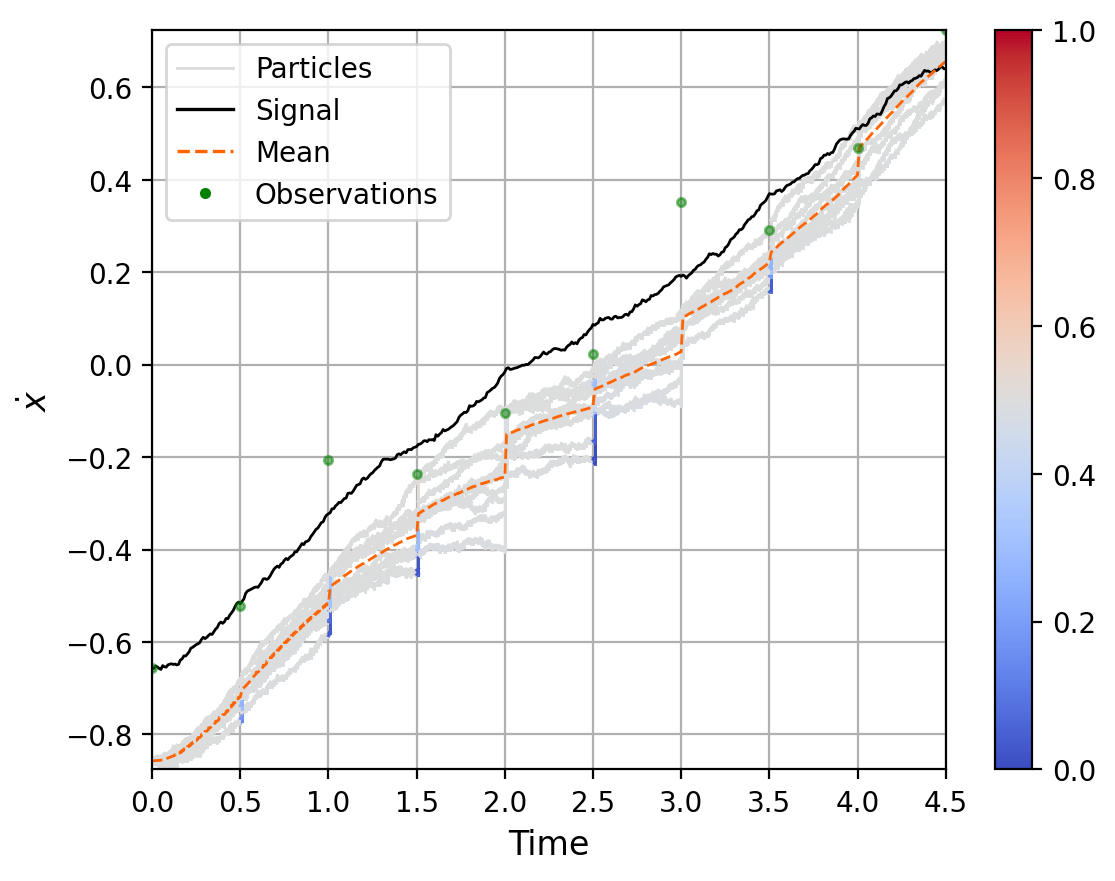}
\caption{
Similar information as in Fig. \ref{figure: deterministic initial condition, split by stable manifold, particle weight history, nPF}, but for the IRnPF. 
One noticeable difference is that with the IRnPF, for each control solution, $\gamma = 5$ advected realizations are produced. 
Hence the generation of the prior distribution has greater resolution at nearly the same computational cost. 
}
\label{figure: deterministic initial condition, split by stable manifold, particle weight history, IRnPF}
\end{figure}

\begin{table}[htbp]
\caption{Monte Carlo Analysis Results; Approximately Best Values for Each Criteria \textcolor{Emerald}{Highlighted}.}
\begin{center}
\begin{tabular}{|c|c|c|c|c|c|}
\hline
\textbf{Experiment}&\multicolumn{5}{|c|}{\textbf{Filtering Method}} \\
\cline{2-6} 
\textbf{Statistic} & $\textrm{PF}_{\textrm{1E1}}$ & $\textrm{PF}_{\textrm{1E2}}$ & $\textrm{PF}_{\textrm{1E3}}$ & $\textrm{nPF}_{\textrm{1E1}}$ & $\textrm{IRnPF}^{\gamma = 10}_{\textrm{1E1}}$ \\
\hline
Avg. RMSE & 0.84 & 0.58 & \textcolor{Emerald}{0.42} & 0.52 & 
\textcolor{Emerald}{0.40} \\
Avg. Min. RMSE & 0.18 & 0.17 & \textcolor{Emerald}{0.16} & 0.17 & 
\textcolor{Emerald}{0.16} \\
Avg. Max. RMSE & 2.25 & 1.57 & \textcolor{Emerald}{0.91} & 1.26 & 
\textcolor{Emerald}{0.87} \\
Avg. (N\textsubscript{eff} / N) & 0.34 & 0.25 & 0.26 & 0.40 & 
\textcolor{Emerald}{0.50} \\
Total Runtime (s) & 4.3 & 19.5 & 178 & 814 & 
1007 \\
\hline
\end{tabular}
\label{table: separated by stable manifold, deterministic initial conditions, monte carlo}
\end{center}
\end{table}

\begin{figure}[hp]
\centering
\includegraphics[scale=0.54]{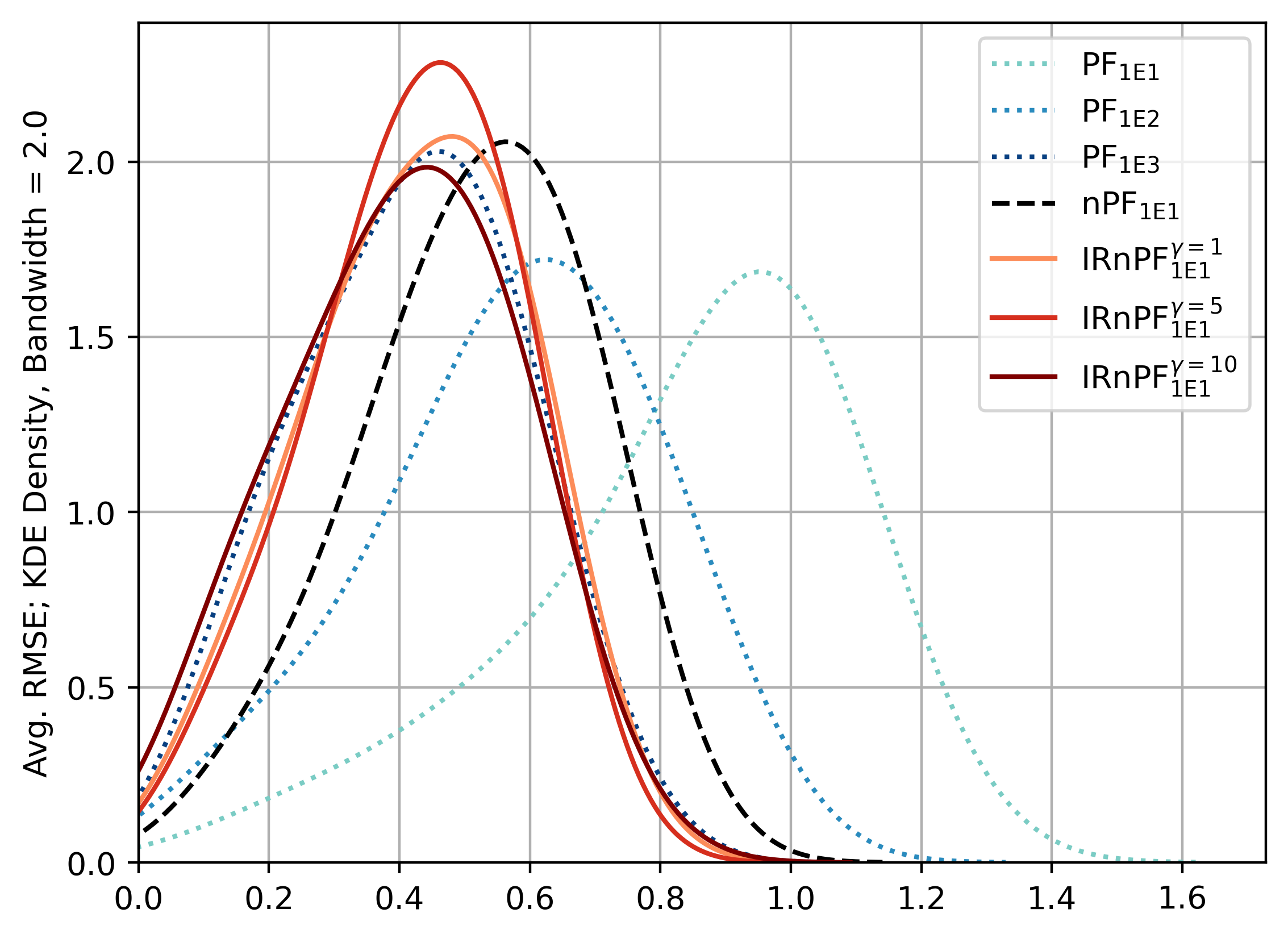}
\caption{
Histogram for Avg. RMSE data shown in Table \ref{table: separated by stable manifold, deterministic initial conditions, monte carlo}.
}
\label{figure: Average RMSE histogram}
\end{figure}

\begin{figure}[hp]
\centering
\includegraphics[scale=0.54]{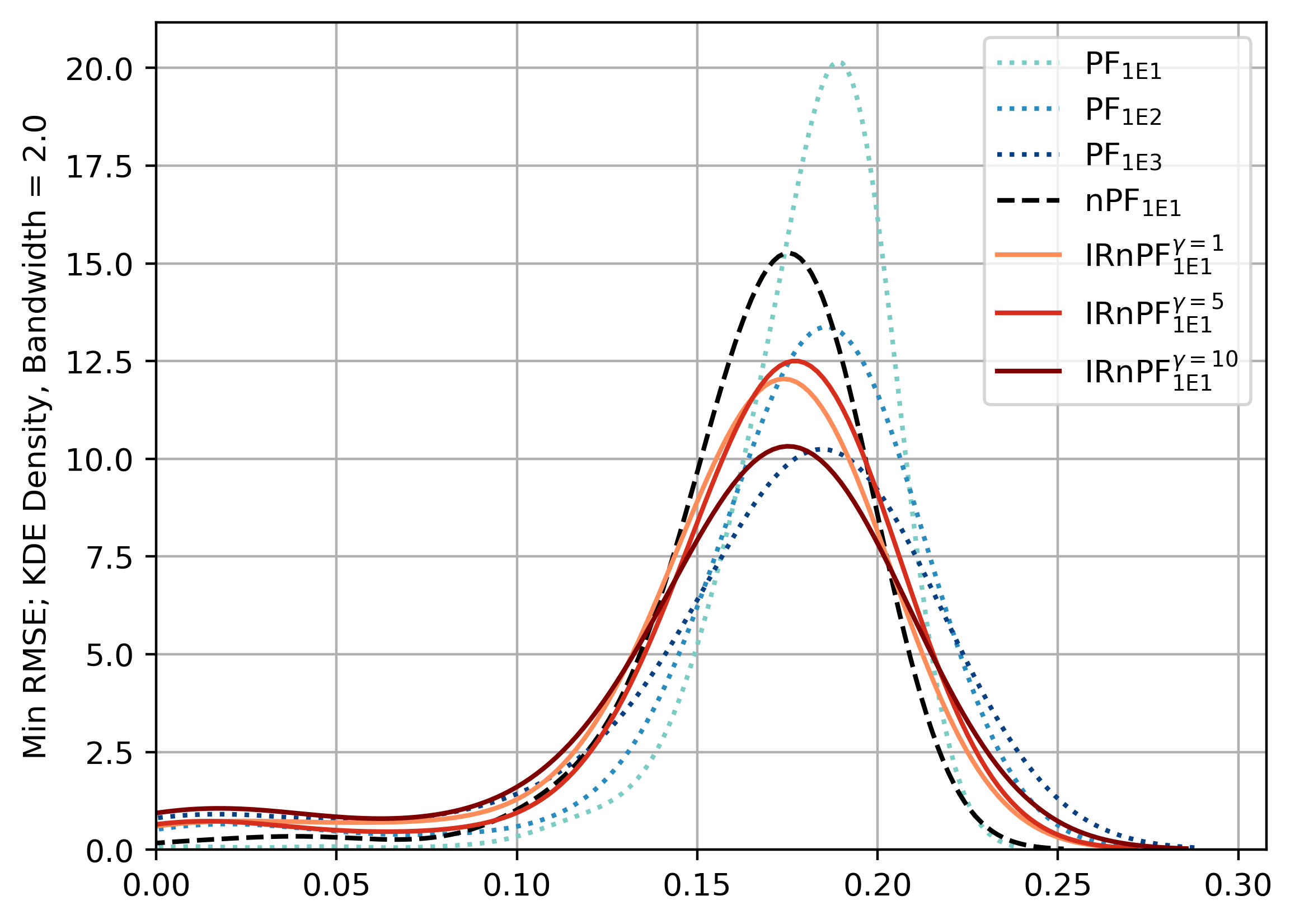}
\caption{
Histogram for Min. RMSE data shown in Table \ref{table: separated by stable manifold, deterministic initial conditions, monte carlo}.
}
\label{figure: minimum RMSE histogram}
\end{figure}

\begin{figure}[hp]
\centering
\includegraphics[scale=0.54]{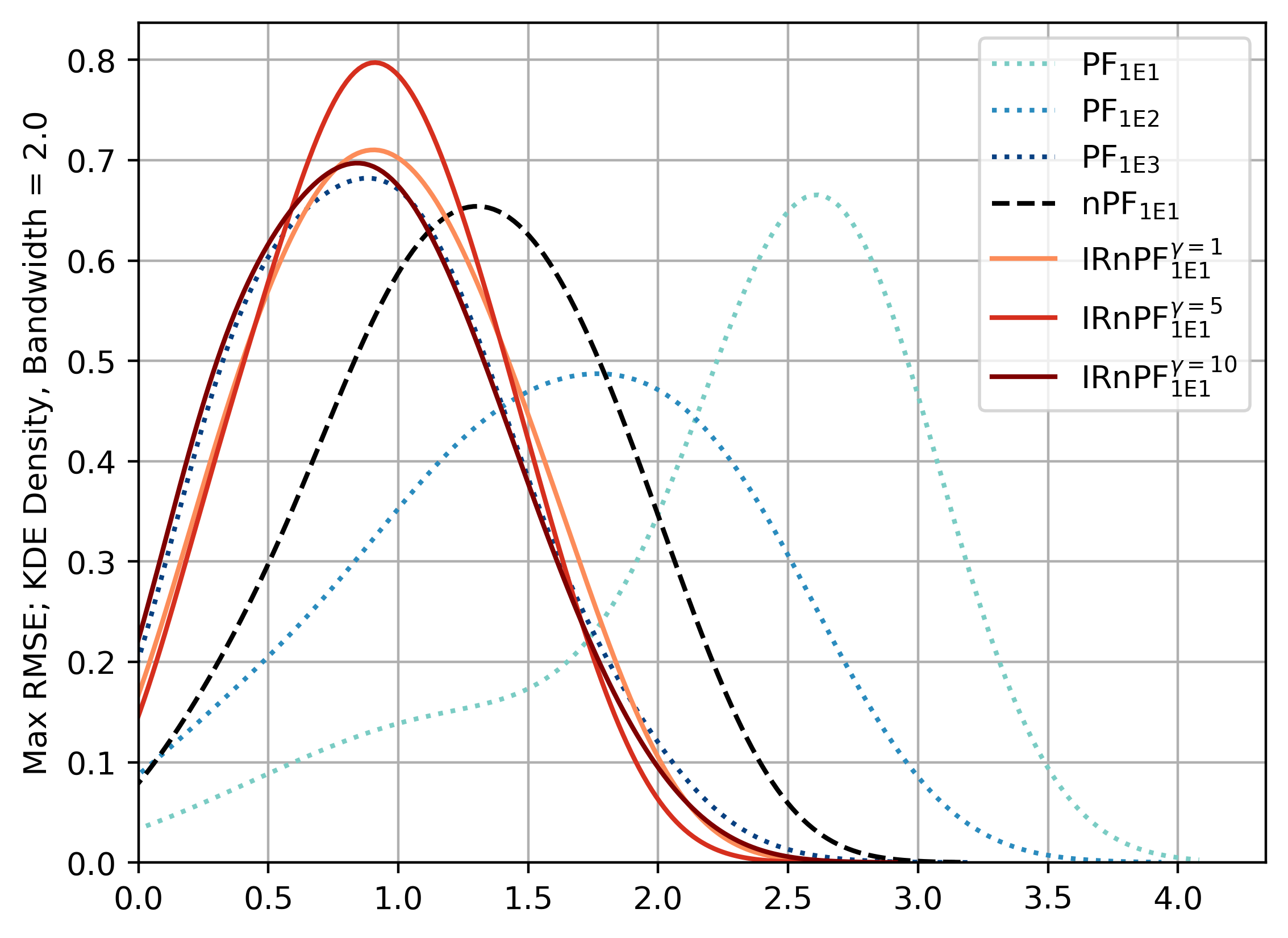}
\caption{
Histogram for Max. RMSE data shown in Table \ref{table: separated by stable manifold, deterministic initial conditions, monte carlo}.
}
\label{figure: maximum RMSE histogram}
\end{figure}

The main results are summarized in Table \ref{table: separated by stable manifold, deterministic initial conditions, monte carlo}. 
We indicate the number of particles used in the PF method with the subscript (e.g., 1E1) and the number of nudged particles for the nPF and IRnPF similarly. 
The superscript for the IRnPF method indicates the multiplicative number of particles advected for each control (see Section \ref{subsection: the IRnPF method}). 
The root mean square error (RMSE) and the Min./Max. RSME for a given simulation realization $\omega$ are defined as, 
\begin{align*}
\op{RMSE}(X_t, p^N_t) (\omega)
\equiv \frac{1}{4.5} \int_0^{4.5} \| X_s(\omega) - \E[p^N_s(\omega)]{x} \|_2 ds, 
\end{align*}
\begin{align*}
\op{Min. RMSE}(X_t, p^N_t) (\omega)
\equiv \min_{s \in [0, 4.5]} \| X_s(\omega) - \E[p^N_s(\omega)]{x} \|_2,
\end{align*}
where $X_s$ is the true signal path for a given realization and $\op{Max. RMSE}$ is defined in an analogous way. 
The Avg. of each of these metrics, as reported in Table \ref{table: separated by stable manifold, deterministic initial conditions, monte carlo}, is the average over the 400 simulations. 
Histograms\footnote{Kernel Density Estimates (KDE)} of the RMSE data in Table \ref{table: separated by stable manifold, deterministic initial conditions, monte carlo} is shown in Figs. \ref{figure: Average RMSE histogram}, \ref{figure: minimum RMSE histogram}, and \ref{figure: maximum RMSE histogram} to provide a greater understanding of the statistics of the MC runs. 

The behavior and performance of the three methods for the example realization described in Section \ref{section: numerical experiments} and shown in Figs. \ref{figure: global view, separated by stable manifold, prediction flow 8, PF} through \ref{figure: deterministic initial condition, split by stable manifold, particle weight history, IRnPF} is apparent in the data of Table \ref{table: separated by stable manifold, deterministic initial conditions, monte carlo}.
The lower Avg. Max. RMSE for the nPF\textsubscript{1E1} versus the PF\textsubscript{1E1} shows that on average the nPF is able to counter the dynamical flow sufficiently long to loiter near the stable manifold before slowly being forced along the unstable direction as shown in Fig. \ref{figure: global view, separated by stable manifold, prediction flow 8, nPF}. 
Given that the initial RMSE is 0.2, the Avg. Min. RMSE for the PF\textsubscript{1E1} case shows that the PF with few particles is largely unable to ever get closer to the signal. 
Increasing the number of particles to the PF\textsubscript{1E2} case enables the standard PF to produce RMSE metrics nearer to that of the nPF\textsubscript{1E1}, but the nPF still has superior effective sample size. 
Of course, the main disadvantage of the nPF\textsubscript{1E1} is the computation complexity due to calculating the control. 
Increasing the number of particles further in the standard approach, the PF\textsubscript{1E3} case is able to outperform the nPF\textsubscript{1E1} in all RMSE criteria with a saving of about 4.5 times the computational effort. 

The new approach proposed in this paper, the IRnPF, is competitive with the PF\textsubscript{1E3} case, even if only 10 particles are used. 
The IRnPF method retains superior effective sample size and has approximately the same computational complexity as the nPF. 
The nudging parameters for the nPF variants (nPF and IRnPF) have not been optimized in these experiments and could potentially be more competitive. 
It is known that the number of particles for the standard PF scales exponentially with the dimension of the problem. 
Therefore there may be a crossover where the nPF and IRnPF methods are more advantageous in terms of RMSE, effective sample size, and computational complexity criteria, as the dimensionality of the problem with similar dynamical features is increased. 

\section{Conclusions}
\label{section: conclusions}

This paper has built on the optimal control approach of the nPF developed in Lingala et al. \cite{Lingala:2014bu}, Yeong et al. \cite{Yeong:2020}, and Beeson and Namachichivaya \cite{Beeson:2020nd} for increasingly chaotic systems, by adding an intermediate optimal deterministic resampling approach using the modified Cram\'{e}r-von Mises distance developed by Hanebeck and Klumpp \cite{Hanebeck:2008.ieee.cmfiis} and Eberhardt et al. \cite{Eberhardt:2010.acc}, to mitigate the isolated intermediate particle degeneracy condition of the nudged method. 
The result is a method that can more continuously apply control to adapt to difficult dynamical regimes such as the separating flow problem of Section \ref{section: numerical experiments} for the stochastic Duffing oscillator with erroneous deterministic initial condition. 
The new intermediate resampling nudged particle filter is consistently able to overcome the separating flow with only a few particles. 

Ongoing improvements by the authors in tuning the nudged component and intermediate resampling may result in the method outperforming the PF, not only in estimation quality, but also in computational runtime, for similarly complex dynamical problems in higher dimensions. 
Improvements in the nudging optimal control theory are also being pursued. 

One theoretical drawback of including the intermediate resampling is the potential loss of independence of the particles; this independence is also potentially lost after Bayes' update if resampling occurs due to the effective sample size decreasing below a user defined threshold. 


\printbibliography
\end{refsection}

\end{document}